\documentclass[journal]{IEEEtran}
 
\usepackage{cite}
\usepackage{amsmath,amssymb,amsfonts}
\usepackage{algorithmic}
\usepackage{algorithm}
\usepackage{graphicx}
\usepackage{textcomp}
\usepackage{xcolor}
\usepackage{booktabs}
\usepackage{multirow}
\usepackage{array}
\usepackage{colortbl}
\usepackage[hidelinks]{hyperref}
\usepackage{url}
\usepackage{tabularx}
\definecolor{headerblue}{RGB}{0, 112, 192}
\definecolor{orange80}{RGB}{255,165,0}
\definecolor{lightblue}{rgb}{0.85, 0.92, 1}
\usepackage{cite}
\usepackage{amsmath,amssymb,amsfonts}
\usepackage{algorithmic}
\usepackage{algorithm}
\usepackage{graphicx}
\usepackage{textcomp}
\usepackage{xcolor}
\usepackage{listings}
\usepackage{booktabs}
\usepackage{multirow}
\usepackage{hyperref}
\usepackage{booktabs}
\usepackage{float}
\usepackage{placeins}
\usepackage[table]{xcolor}
\definecolor{bestgreen}{RGB}{198,239,206}

\def\BibTeX{{\rm B\kern-.05em{\sc i\kern-.025em b}\kern-.08em
    T\kern-.1667em\lower.7ex\hbox{E}\kern-.125emX}}

\ifCLASSINFOpdf
\else
\fi
\begin{document}
%
\title{FeedbackLLM: Metadata driven Multi-Agentic Language Agnostic Test Case Generator with Evolving prompt and Coverage Feedback}
%
%
%

\author{Kushal Jasti,~\IEEEmembership{Department of Computer Science and Engineering, SRM University AP}\\
        Tejamani Prashanth Sahu,~\IEEEmembership{Department of Computer Science and Engineering, SRM University AP}\\
         Rishitha Pentyala,~\IEEEmembership{Department of Computer Science and Engineering, SRM University AP}
        Muvvala Mohit,~\IEEEmembership{Department of Computer Science and Engineering, SRM University AP}\\
        and~Vivek~Yelleti,~\IEEEmembership{Department of Computer Science and Engineering, SRM University AP}
\thanks{All authors are with the Department
of Computer Science and Engineering, SRM University AP, India.}
\thanks{Corresponding Author: Vivek Yelleti, was with the Department
of Computer Science and Engineering, SRM University AP,
India, e-mail: vivek.yelleti@gmail.com.}
}

%
%

\markboth{Journal of \LaTeX\ Class Files,~Vol.~14, No.~8, August~2015}%
{Shell \MakeLowercase{\textit{et al.}}: Bare Demo of IEEEtran.cls for IEEE Journals}
%



\maketitle

\begin{abstract}
Traditional approaches to test case generation often involve manual effort and incur significant computational overhead. Additionally, these approaches are not scalable, and hence, unsuitable for complex software systems. Recently, Large Language Models (LLMs) have been applied to software testing. However, single-shot prompt engineering-based approaches tend to hallucinate and generate redundant test cases, resulting in fewer branches. To handle the above-mentioned limitations, in this paper, we propose FeedbackLLM, a novel automated language-agnostic test case generation framework based on tightly coupled two-stage approach. In the first stage, FeedbackLLM extracts the input constraints by parsing source code and generates the possible test cases. The quality of the test cases is evaluated in the second stage by the following two specialized LLM feedback agents: (i) \textit{Line Feedback Agent}: extracts the metadata related to missed line executions and (ii) \textit{Branch Feedback Agent}: extracts the metadata of the unexecuted branch conditions. The above agents operate in a two-stage process, communicating in tandem, and this procedure is repeated for $k$-steps. Further, we also introduced a redundancy prevention cache to avoid duplicate API requests and avoid unnecessary execution cycles. The performance of the proposed architecture is evaluated on the standard benchmark programs related to C and Python programs. FeedbackLLM demonstrated more line and branch coverage than baseline tools while scaling linearly in execution time.
\end{abstract}


\section{Introduction}
Software Testing is referred to as a systematic process of trying to understand the behavior of the software in different situations. This is one of the fundamental and important stages in the software development life cycle (SDLC). This involves high resource-intensive tasks since it involves testing the entire software. As and when the code bases become larger and more complex, it becomes exponentially more difficult to yield higher testing metrics such as line coverage and branch coverage. Hence, this stage occupied about 10-20 \% of the total budget involved in the software product development. The end goal of software testing is to find edge cases and subtle bugs in the underlying software program before it is put into production. 

To address the above limitations, researchers paved the path for automated software testing, where specialized software is developed to analyze the code bases. In the testing, the automation could be achieved in the following ways: 
\begin{enumerate}
    \item the process of performing the testing on the underlying software could be automated;
    \item the process of generating the test cases could be automated;
    \item both the test case generation and testing the software could be automated
\end{enumerate}
The following are the major benefits of the automated software testing: (i) less to no human intervention; (ii) enhanced quality assurance; (iii) testing frequency could be increased; (iv) smoothens the continuous integration (CI)/ continuous development (CD) pipelines. Among these the generation of test cases is highly critical because of the quality of the test cases decides whether the software is perfectly analyzed or not. Poor test cases ultimately lead to poor quality assurance over the software. The industry standard for finding vulnerabilities and checking formal logic has been fuzz testing frameworks like AFL++ and bounded model checkers like CBMC. But these traditional tools are not without limitations, even with their extensive adoption. Fuzzers depend heavily on random mutations, which can be inefficient at solving complex branch constraints (e.g., particular string comparisons or cryptographic checks). On the other hand, bounded model checkers use symbolic execution and loop unwinding. This leads to state-space explosion and exponential time complexity for analyzing large programs. Generating diverse and high-quality test suites often requires computationally expensive procedures that can take hours or even days.

Large Language Models (LLMs) have revolutionized several phases of the SDLC. LLMs are neural network architectures that mimic the thinking capabilities and learning patterns of humans. They are increasingly used for code synthesis by integrating it as an API, generating the code based on the given scenario or logic, software bug detection, control flow graph generation, etc. Big tech companies like Meta, Google, etc. are rigorously working towards the LLM-driven automation. Due to these reasons, we are witnessing a swift progress in the field that opens a paradigm shift in software analysis. As explained earlier, the reasoning skills of LLMs enhanced the semantic understanding of programming syntax, logic flows, and edge cases. Recently, some of the  studies have shown that LLMs \cite{10.1145/3802827}, \cite{baqar2025future} are capable of generating test cases dynamically based on source code structures. However, the following are the major limitations of the extant LLM-based test generators: 
\begin{enumerate}
    \item They rely on zero-shot or single-agent which ultimately limits the exploration capabilities; and
    \item Extant approaches are mainly language-specific or language-dependent approaches. Hence, they could not be applied to multiple programming languages. 
\end{enumerate}

The above limitations motivated us to propose \textbf{AutoTCG}, a $k$-step language agnostic Test Case Generator. AutoTCG employs the evolving prompt in their two-stage agentic pipeline, where base prompts receive the metadata of the unexecuted code lines and branches for information, which helps the models to diversify the test case generation process. The following is the procedure employed in the tightly coupled two-stage agentic pipeline: (i) In the first stage, LLM analyzes the given complex code and generates the test cases as per the base prompt. (ii) In the second stage, it invokes the Line feedback agent and Branch feedback agent to derive the unexecuted line and branch information. These agents generate targeted constraints which are automatically incorporated into the generation prompt for the next iteration. Once the meta data is extracted, then it is be passed to the prompt of the agent employed in the first stage. Additionally, to ensure API efficiency and avoid redundant executions, AutoTCG employs the Redundancy Prevention Cache as used in the kS-LLM approach that stores previously generated inputs and filters duplicates in a native manner.

The following are the main contributions in this paper:
\begin{itemize} 
    \item We proposed AutoTCG on a fully automated language-agnostic  framework, which could work with multiple languages.
    \item We introduced Multi-Agent Feedback Loop that derives the metadata of the unexecuted lines and branch code to enhance line coverage and branch coverage.
    \item We introduced the Redundancy Prevention Cache to decrease API quota usage and execution time, resulting in linear scalability.
\end{itemize}

The remaining sections of the paper are organized as follows. Section II surveys related work. III. AutoTCG Architecture and Approach In Section III, we describe the architecture and approach of AutoTCG. Section IV concerns the experimental setup and Section V discusses the results and analysis. Threats to validity are discussed in Section VI. Section VII concludes the paper.
\section{Related Works}
Algarsamy et al. \cite{ALAGARSAMY2024107565} generated the unit test cases using the assertion approach and named it as A3Test. Also, they added the ability to include domain knowledge in the test case generation process. Ni et al. \cite{11229491} proposed CasModaTest which uses a tiered technique for generating the test cases in the following manner: (i) the given code is first split into several cascaded ones; (ii) test prefix generation is then called; and (iii) test oracle generation is then called. Arya et al. \cite{11264985} suggested a method for generating the unit test cases based on NLP and reinforcement learning. Kikuma et al. \cite{10.1145/3368926.3369679} suggested a machine learning-based approach to extract homogenous test cases which are completely independent of abilities and mimic the writing style of an expert who needs to write the test cases. Its approach produces homogeneous test cases based on the requirements specification document and previous development knowledge. Adversarial test scenarios to validate the efficacy of autonomous vehicles were proposed by Tuncali et al. \cite{8500421}. Tuncali et al. \cite{8911483} in another paper proposed the usage of a software requirements document to create the test cases to be used as adversarial test cases. Esnaashari et al. \cite{esnaashari2021automation} presented a memetic algorithm in which the genetic algorithm generates the test cases. However, they developed reinforcement learning as the local search technique to improve the exploitation and to keep the variety. Lafi et al. \cite{9491761} used machine learning techniques to generate test cases based on the analysis of the given software requirements document. Deeprest \cite{10.1145/3691620.3695511} by Corradini et al. used a reinforcement learning approach to generate test cases for Rest APIs. The branch execution is analyzed, coded and the data is extracted by PySE, which was proposed by Koo et al. \cite{8730198}. This policy is then constantly adjusted, using data from the underlying reinforcement learning agent. Tufano et al. \cite{tufano2020unit} proposed unit test case generation based on transformers. They also boosted performance by embedding the focal context in the learning process. By this approach, the transformer could complete more connected unit test cases by giving more weight to the appropriate token. Samuel et al. \cite{samuel2008automatic} proposed the technique for generating test cases by looking into the UML diagrams. Liu et al. \cite{10.1145/3575693.3575707} presented the NNSmith tool. First, it analyzes deep learning compilers using the fuzzing approach and then generates test cases. Their innovative fuzzing technique contributed to the creation of the various and reliable test scenarios. Yue et al. \cite{10.1145/2771783.2771799} built a Toucan4Test tool which considers the specification of the testing team and analyses these documents to generate the test cases. Seqerloo et al. \cite{yazdani2019automatic} analyzed 30 automatically generated TCSs and validated their method in the two industrial case studies. They thought business model analysis to create test cases. They increased the diversity and relevance to the business challenge by converting the business model to state graphs and using these to create the test cases. Steenhoek et al. \cite{11026897} suggest a two-stage process where the test cases are generated by LLMs. Then, the reinforcement learning technique is applied to reduce the test smells in the produced test cases by tuning five coding quality controls. Sami et al. \cite{sami2024tool} proposed LLM based on prompt engineering for generating test cases. But to do this it examines the requirements document. Mathur et al. \cite{10112971} generated the test scenarios using T5 and GPT-3. These two approaches use context-based meaning to find relevant keywords and generate test cases based on keywords. Zhang et al. \cite{zhang2024testbench} evaluated LLM models like CodeLlama-13b, GPT-3.5, and GPT-4 for generating the class-level test cases. They found that the smaller models are contained within the full context and tend to be more sensitive to the noise. 

Yang et al. \cite{yang2025llm} proposed a three-phase approach for generating test cases for dynamic programming languages such as Python. The three steps they use are: (i) first, they check the type of the parameter; (ii) then, they collect important data during mutations; and (iii) then, they use LLMs to fix the generated test cases. ChatUniTest is a generate-validate-repair approach proposed by Chen et al. \cite{10.1145/3663529.3663801}. We use LLMs here for those three processes separately, and also to generate unit test cases. Hasan et al. \cite{11025799} presented an approach to align the test case generation to business needs. They tested a number of models and found the Llama and Mistral models to be better than the rest. Kurian \cite{10172742} developed an LLM-based design for Scade, a modern programming language used in safety-critical systems like railways. Lukasczyk and Fraser have suggested a fast system of independent agents \cite{10.1145/3510454.3516829}. Xue et al. \cite{xue2024llm4fin} constructed a domain-specific framework, named LLM4Fin, by fine-tuning the pre-trained model. This model was used to create the test cases. The model’s performance was verified by real world stock trading. In their empirical investigation, their LLM4Fin outperformed both ChatGPT and other LLM models. Korraplu et al. \cite{korraprolu2025test} performed a benchmark analysis on six popular LLMs. In a number of categories their results showed the Gemini came out on top. Deng et al. \cite{deng2024llm} proposed a method to generate the prompts by considering and analyzing the register transfer level behavior. The performance of the LLM model depends on the quality of the prompt. Gao et al. \cite{gao2025prompt} proposed an ideal way to generate user-defined prompts for automated test case generation. The method suggested by Schafer et al. \cite{10329992} is that the user uploads the API of the project first. Then the prompts are created automatically based on the signature and function type as per the developer’s guidance. Koziolek et al. \cite{10711016} used search-based methods and symbolic execution to generate test cases. The performance of their model is validated using the control logics of PLC and DLC .

\section{Proposed Approach}
In this section, we present the architecture and internal mechanics of FeedbackLLM. We start the section with the overview of the proposed approach and then discuss each inner component effectively. The flowchart and block diagram of the proposed FeedbackLLM are presented in Fig. 1 and Fig. 2, respectively. The step-by-step of the proposed approach is also presented in Algorithm 1.

\begin{algorithm*}[ht]
\caption{FeedbackLLM Iterative Feedback Loop}
\begin{algorithmic}[1]
\REQUIRE Source code $S$, coverage threshold $T=90\%$, maximum iterations $k_{\max}=10$
\STATE $cache\leftarrow\emptyset$
\STATE $prompt\_refinements\leftarrow\emptyset$
\STATE $k\leftarrow0$
\WHILE{$k<k_{\max}$}
    \STATE $prompt\leftarrow\text{Merge}(\text{Baseline},prompt\_refinements)$
    \STATE $json\_tests\leftarrow\text{CallLLM}(prompt,cache)$
    \STATE $new\_tests\leftarrow\text{ParseAndFilter}(json\_tests,cache)$
    \STATE $cache\leftarrow cache\cup new\_tests$
    \STATE $\text{Execute}(S,cache)$
    \STATE $C_{line},C_{branch},Gaps\leftarrow\text{EvaluateCoverage}()$
    \STATE $C_{total}\leftarrow\frac{C_{line}+C_{branch}}{2}$
    \IF{$C_{total}\geq T$}
        \STATE \textbf{break}
    \ENDIF
    \STATE $feedback_{line}\leftarrow\text{LineFeedbackAgent}(S,Gaps_{line},prompt)$
    \STATE $feedback_{branch}\leftarrow\text{BranchFeedbackAgent}(S,Gaps_{branch},prompt)$
    \STATE $prompt\_refinements\leftarrow feedback_{line}\cup feedback_{branch}$
    \STATE $k\leftarrow k+1$
\ENDWHILE
\RETURN $cache,C_{line},C_{branch}$
\end{algorithmic}
\end{algorithm*}

\subsection{Overview of the proposed approach}
FeedbackLLM is a tightly coupled two-stage agentic pipeline where the following stages are invoked in tandem with each other. This entire pipeline uses a prompt-engineering-based method, and we arrived at these prompts through extensive experimentation. In the first stage, LLM is employed, which acts as an agent and invokes the base prompt as provided in Listing 1. It analyzes the given complex code and generates the test cases as per the base prompt. In the second stage, we employed two individual LLMs, which are treated as a Line feedback agent and a Branch feedback agent. These agents are specifically employed to derive the information on not-executed lines and branches. These agents generate targeted constraints that are automatically incorporated into the generation prompt for the next iteration. This completes the first iteration, and from the second iteration onwards, along with the base prompt, the metadata of the not executed lines and branches is also concatenated to the base prompt. Additionally, to ensure API efficiency and avoid redundant executions, FeedbackLLM employs the Redundancy Prevention Cache as used in the kS-LLM approach that stores previously generated inputs and filters duplicates in a native manner.
\begin{figure}[htbp]
\centering
\includegraphics[width=1.0\linewidth,height=9cm,keepaspectratio]{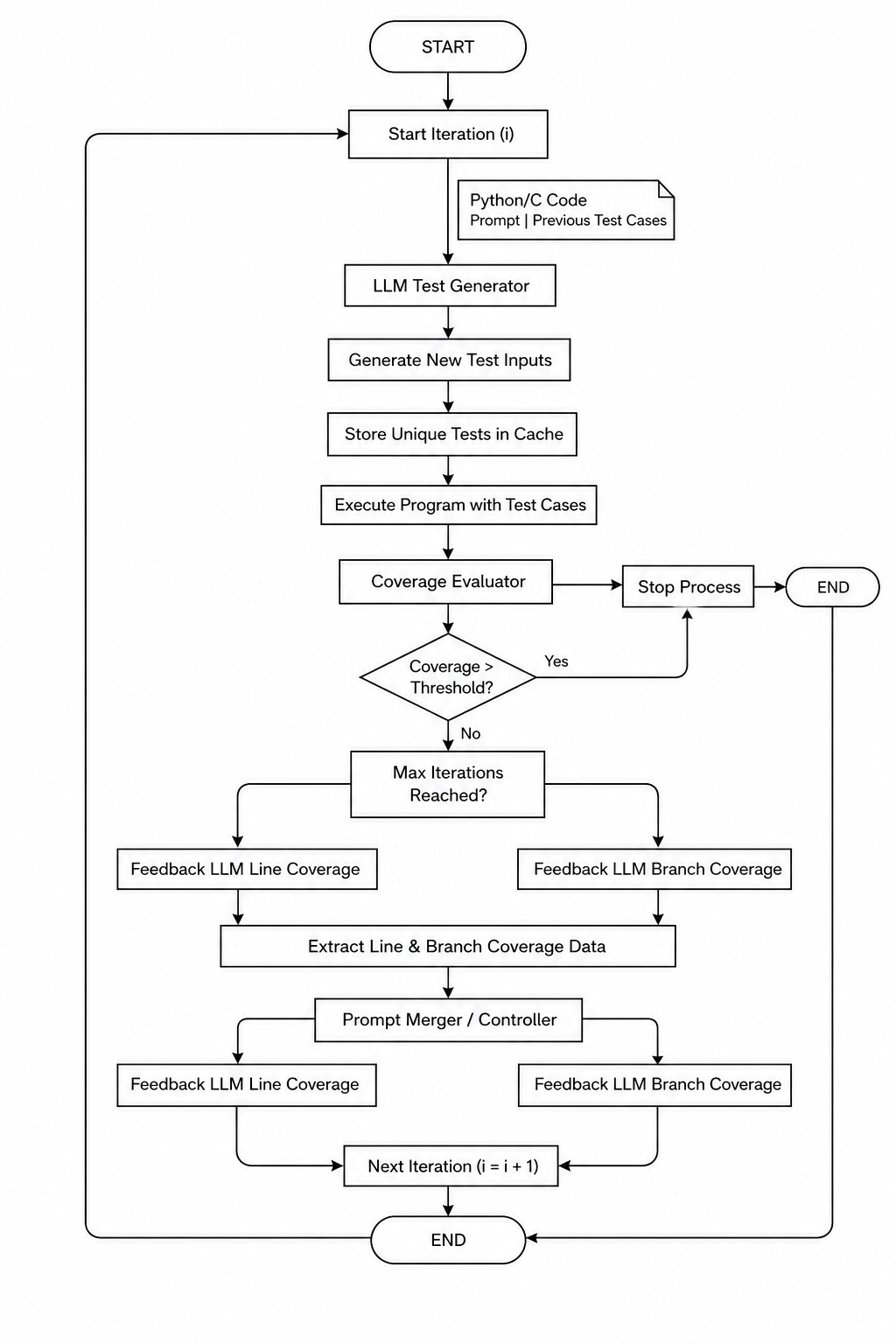}
\caption{Systematic representation of the KSLLM automated test case generation framework.}
\label{fig:architecture}
\end{figure}
\begin{figure*}[!t]
\centering
\includegraphics[width=1.3\columnwidth]{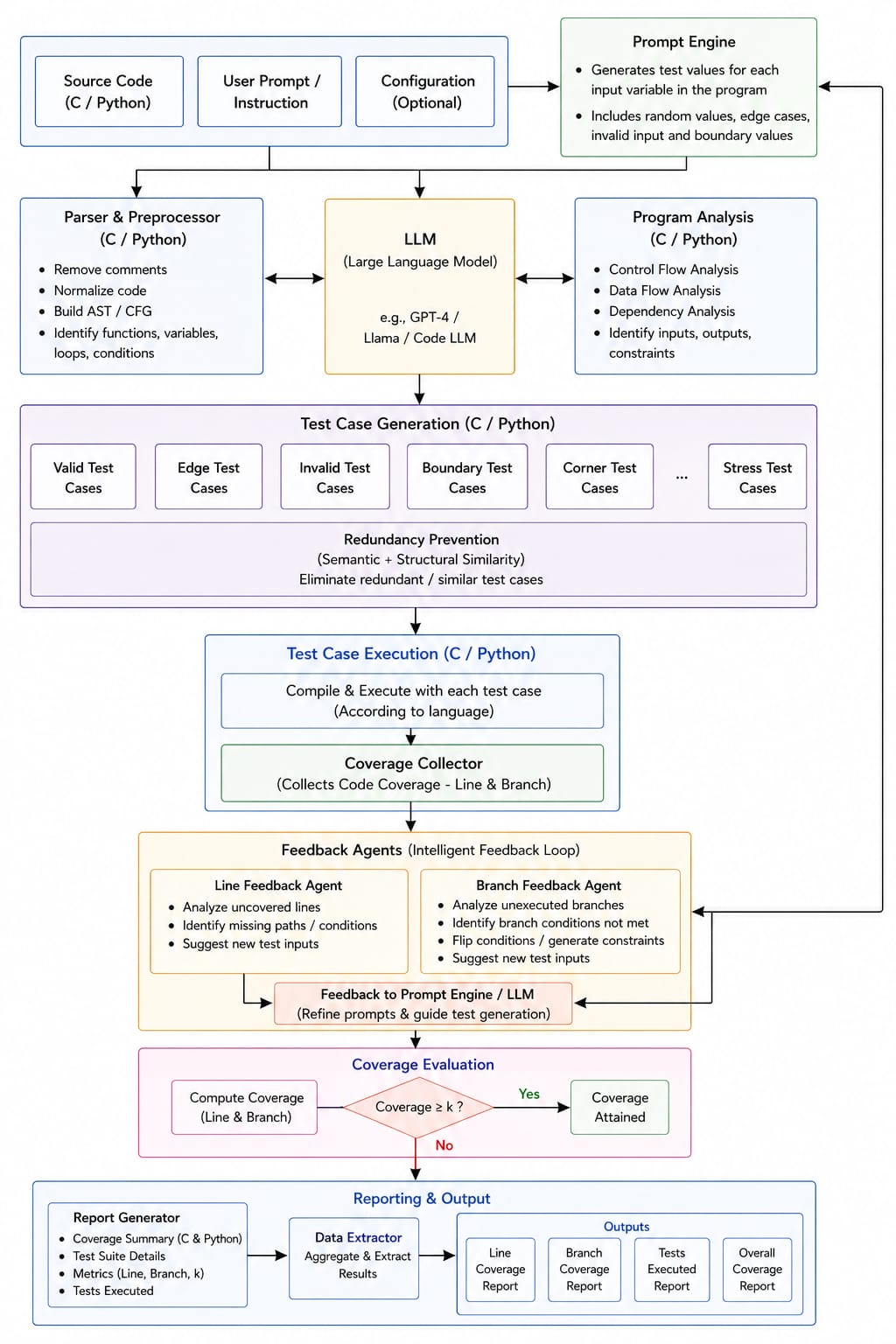}
\caption{Schematic diagram of the proposed approach}
\label{fig:your_label}
\end{figure*}

\lstset{caption={Template of the Baseline LLM Prompt},label={lst:prompt}}
\begin{lstlisting}[language=Python]
Generate diverse test values for the following program.
IMPORTANT: This program calls input exactly {n} times.
Detailed input types (in order):
  Input #1: expects {type_1}
  ...
Include diverse cases: boundary chars, negative integers, large extremes.
Output format must be a JSON object with a "test_cases" key containing a list of lists.
Previously generated values: {cache}
\end{lstlisting}
\subsection{LLM Agent Configuration}
It is important to emphasize that the FeedbackLLM framework utilizes multi-agent architecture, where each functional agent is powered by a Large Language Model (LLM). For this implementation, we used the Google Gemini-2.5-Flash model for all agents, configured via the official Google GenAI SDK.

We employed the following three specialized agents:

\textbf{1) Test Case Generator:}  
It generates various input test cases from the extracted program constraints. This agent uses the Gemini-2.5-Flash model to analyze the structured baseline prompt and generate several candidate test inputs in JSON format. The model was trained to maximize diversity, boundary value exploration and consistency of structured outputs. 

\textbf{2) Line Feedback Agent:}  
The Line Feedback Agent deals with statement level coverage gaps. This agent uses the Gemini-2.5-Flash model to parse the missing line numbers provided by the coverage evaluator and deduces the logical conditions that should be met for executing these line numbers. The agent produces refined prompt instructions that steer subsequent test generation towards unexplored execution paths.

\textbf{3) Branch Feedback Agent:}  
The Branch Feedback Agent is concerned with branch-level coverage gaps. It uses the Gemini-2.5-Flash model to examine partially executed conditionals and infer additional input values needed to generate alternate branch results. The structured feedback obtained is integrated into the prompt refinement pipeline.

All agents use the same LLM configuration to ensure the same reasoning behavior in both Python and C programs. This consistency removes model-specific bias and makes sure performance gains are due to the proposed multi-agent design and not the differences in the underlying LLMs’ capabilities.

\subsection{Multi-Agent Feedback Loop}
Either when the threshold is less than $< 90\%$, or it invokes fewer than $k$ iterations, the framework enters the refinement stage, a core part of the iterative optimization strategy of FeedbackLLM. Unlike traditional single-agent approaches that depend on coarse or aggregated feedback, FeedbackLLM explicitly decomposes the feedback process into specific analytical roles, reducing the cognitive load on any individual model and improving the accuracy of generated test inputs. 

\textbf{Line Feedback Agent:} The Line Feedback Agent works only on statement-level coverage holes. It needs the following three main inputs: (i) the full source code, (ii) the current prompt strategy for test generation, and (iii) a structured list of missing/unexecuted line numbers obtained from the coverage evaluator. From this information, the Line Feedback Agent performs semantic reasoning to explain why some lines were not executed. Such reasons can be the unfulfilled conditional constraints, the lack of diversity in input or the boundary values are not covered . The model then generates a structured JSON response that includes:
\begin{itemize}
    \item A concise explanation of the coverage gap
    \item Identification of input patterns required to trigger the missing lines
    \item Explicit prompt refinements that guide the generator toward these patterns
\end{itemize}
This particular feedback guarantees to steer subsequent iterations towards new execution paths, and not to reproduce old ones that do not work.

\textbf{Branch Feedback Agent:}
In parallel, the Branch Feedback Agent is in charge of analyzing coverage deficiencies at branch level. In particular, it deals with those conditional statements that have been executed and only one of the outcomes (true or false) has been taken. The program code, the list of the partially covered branches and the contextual execution feedback are provided to the Branch Feedback Agent. Then it infers the logical conditions governing each branch and determines the input constraints required to flip the branch outcome. If the condition evaluates only to true, then the Branch Feedback Agent computes the complementary input values to force a false evaluation. Like the Line Feedback Agent it returns a structured JSON with the following:
\begin{itemize}
    \item Analysis of unresolved branch conditions
    \item Required logical constraints for alternative execution paths
    \item Prompt-level modifications to enforce exploration of these paths
\end{itemize}

\textbf{Prompt Merger and Iterative Refinement:}
The output of the Line Feedback Agent and the Branch Feedback Agent are then fed into a deterministic Prompt Merger module. This module merges both JSON responses programmatically into a single ``Additional Focus Areas'' section, coherently merging line-level and branch-level insights. The merged prompt is appended to the baseline generation prompt for the next iteration ($k+1$), thus steering the LLM towards uncharted areas of the program’s execution space.

Formally, the refinement step can be expressed as:
\[
P_{k+1} = P_{base} \cup F_{line} \cup F_{branch}
\]
where $P_{base}$ represents the baseline prompt, and $F_{line}$ and $F_{branch}$ denote the feedback generated by the Line Feedback Agent and the Branch Feedback Agent, respectively.

The multi-agent decomposition allows FeedbackLLM to explore the input space in a fine-grained and targeted manner, which significantly improves the coverage convergence rates while avoiding redundant or ineffective test case generation.

\subsection{Redundancy Prevention Cache (RPC)}
A common problem in iterative LLM generation is duplicate responses generated, which is wasteful in terms of execution time and API quotas. To remedy this, FeedbackLLM has a Redundancy Prevention Cache with $O(1)$ lookup time.

When the LLM produces its JSON payload, FeedbackLLM parses the input lists, casts them to string tuples and then attempts to add them to a Python \texttt{set()}. If the tuple is already there, it is natively ignored. Only novel test cases are written to the `TestCases/` physical directory. The cache contents are returned to the LLM in later prompts (see Listing \ref{lst:prompt}) to guide the model away from explored state spaces.

\subsection{Coverage Evaluation Engine}
The Coverage Evaluator is responsible for compiling, executing, and quantifying the efficacy of the generated test suite.
\begin{itemize}
    \item \textbf{Python Execution:} The framework utilizes the \texttt{coverage.py} module. Test cases are fed via standard input to \texttt{coverage run -a --branch}.
    \item \textbf{C Execution:} The framework compiles the target file natively using \texttt{gcc -fprofile-arcs -ftest-coverage}. The resulting binary is executed, and \texttt{gcov} is invoked to extract execution paths.
\end{itemize}
Following execution, FeedbackLLM generates a JSON artifact detailing \texttt{executed\_lines}, \texttt{missing\_lines}, and \texttt{missing\_branches}. Custom parsing logic extracts the precise Line Coverage and Branch Coverage percentages.

\section{Experimental Setup}
To evaluate the effectiveness of FeedbackLLM, we formulated the following Research Questions (RQs):
\begin{itemize}
    \item \textbf{RQ1:} How does FeedbackLLM compare against zero-shot LLM generation and traditional tools in achieving line and branch coverage?
    \item \textbf{RQ2:} Does the Multi-Agent feedback loop (Line Feedback Agent and Branch Feedback Agent) perform measurably better than a single-agent feedback loop in traversing complex logic paths?
    \item \textbf{RQ3:} How effectively does the Redundancy Prevention Cache scale execution time relative to program complexity?
\end{itemize}

\subsection{Benchmark Dataset}
The framework was evaluated against a suite of 20 standard C programs selected from the PALS and RERS benchmarks, varying in bounds and complexity. Additionally, 20 Python programs containing nested conditionals, loops, and mathematical constraints were utilized to test cross-language applicability.

\subsection{Execution Environment}
All experiments were conducted on a machine running macOS, utilising Python 3.9. The LLM endpoint utilised was Google's \texttt{gemini-2.5-flash} via the official GenAI SDK. For C compilation, standard \texttt{gcc} and \texttt{gcov} utilities were employed.

\begin{figure}[!htbp]
\centering
\includegraphics[width=\columnwidth]{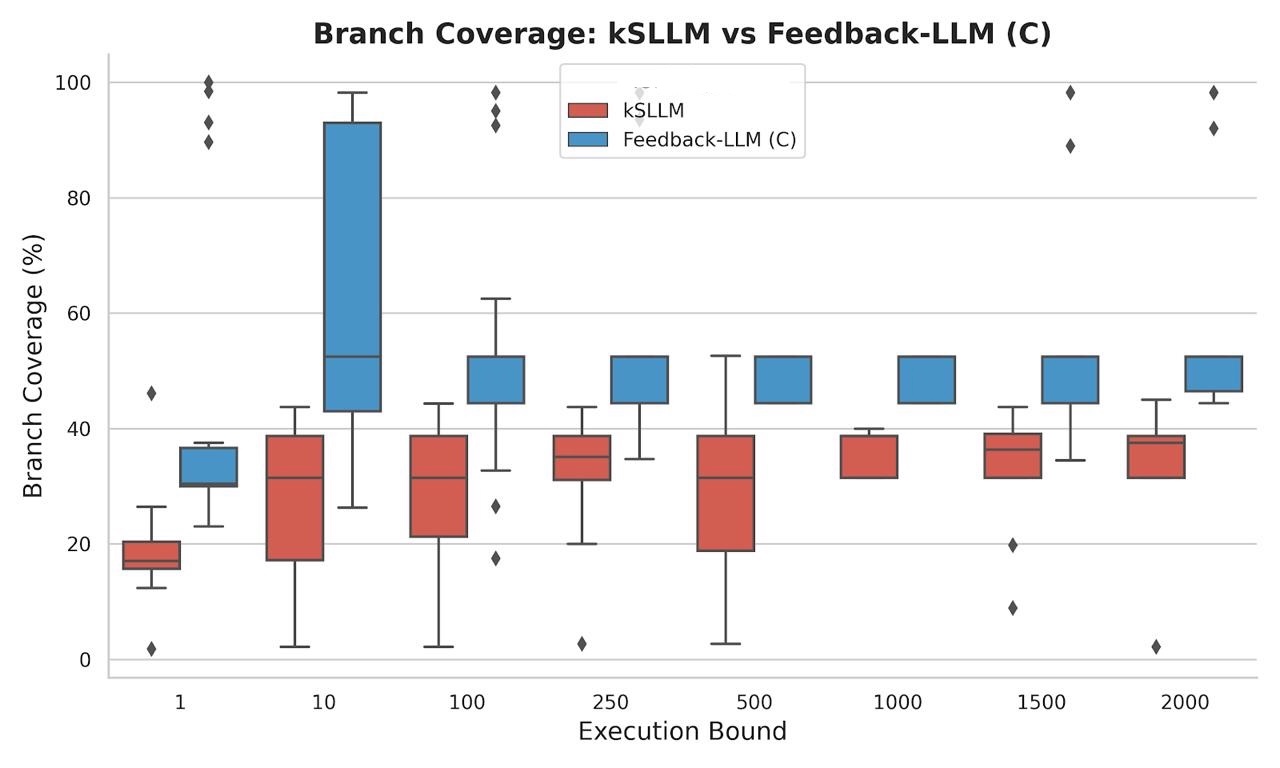}
\caption{Branch Coverage yielded by kS-LLM and FeedbackLLM (C)}
\label{fig:your_label}
\end{figure}

\begin{figure}[!htbp]
\centering
\includegraphics[width=\columnwidth]{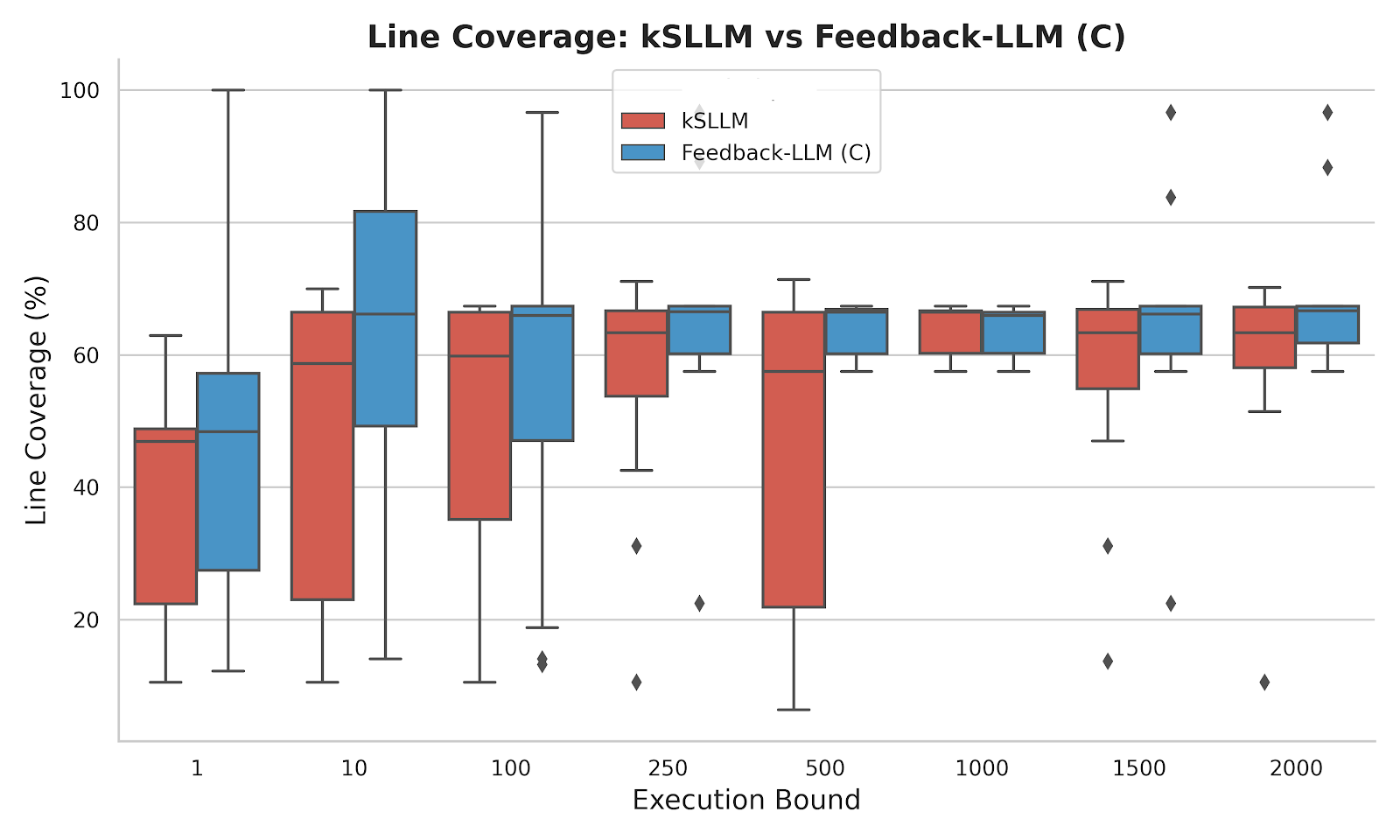}
\caption{Line Coverage yielded by kS-LLM and FeedbackLLM (C) }
\label{fig:your_label}
\end{figure}

\begin{figure}[!htbp]
\centering
\includegraphics[width=\columnwidth]{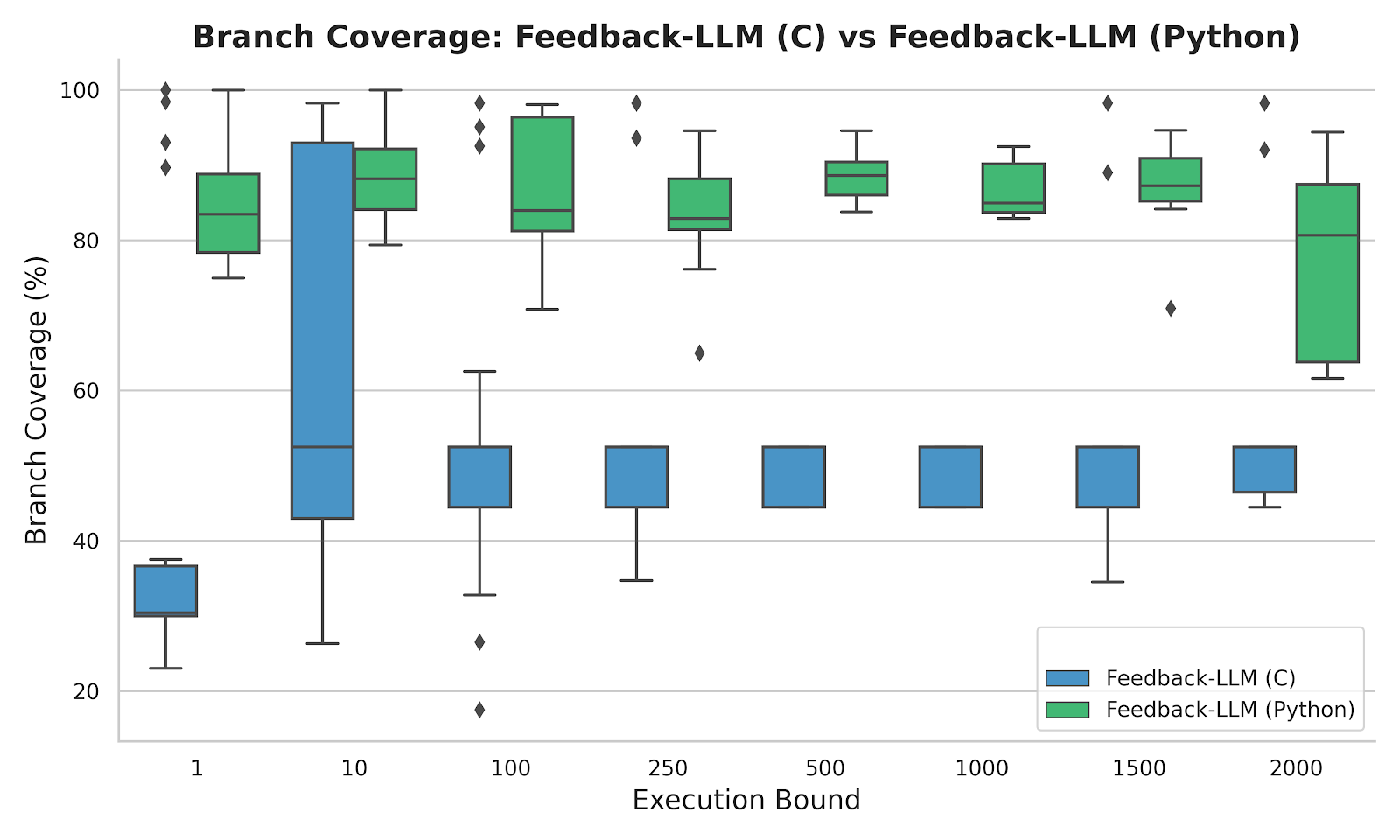}
\caption{Branch Coverage yielded by FeedbackLLM (C) and FeedbackLLM (Py) }
\label{fig:your_label}
\end{figure}

\begin{figure}[!htbp]
\centering
\includegraphics[width=\columnwidth]{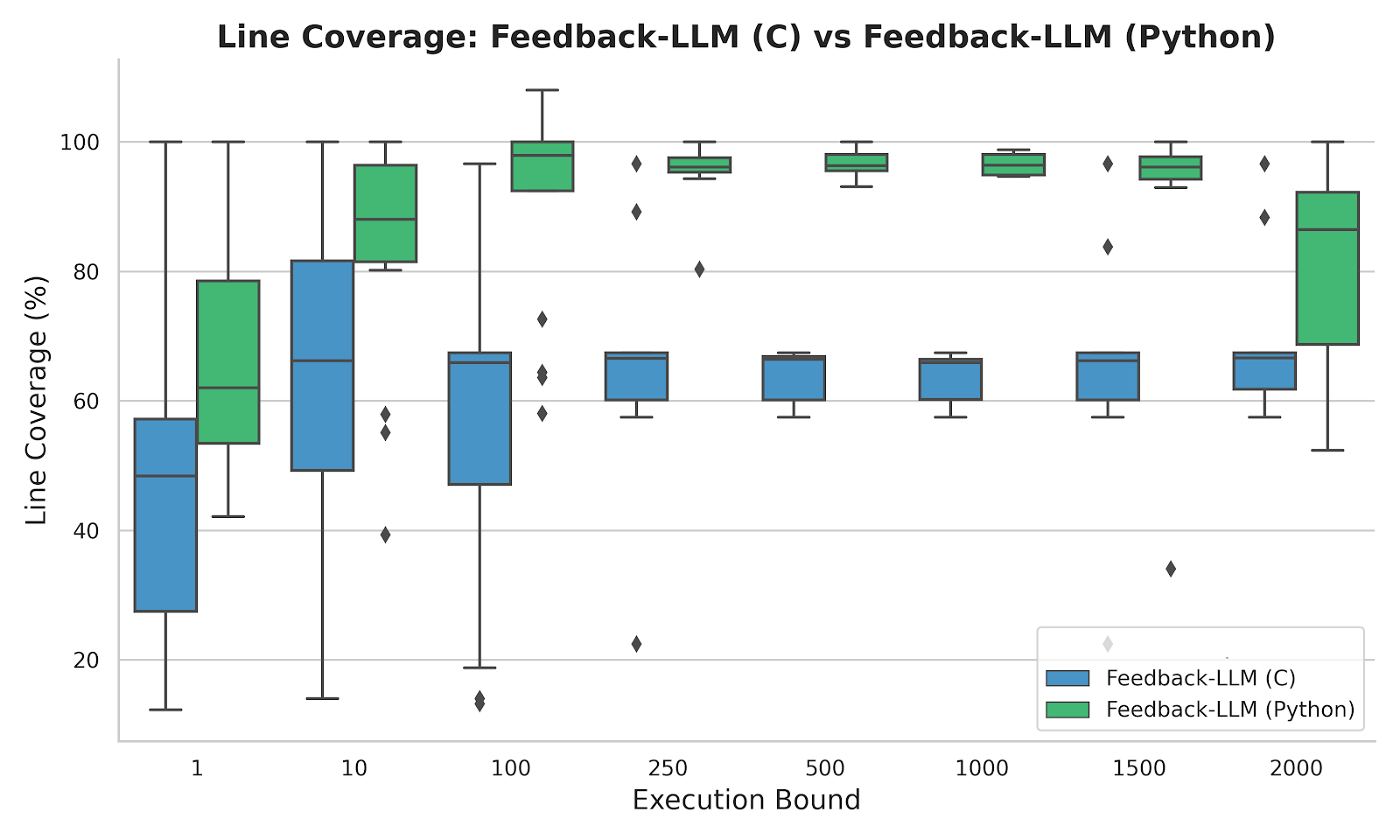}
\caption{Line Coverage yielded by FeedbackLLM (C) and FeedbackLLM (Py) }
\label{fig:your_label}
\end{figure}

\begin{table*}[!htbp]
\centering
\scriptsize
\caption{Coverage and Execution Time Results of KSLLM (Bounds 1--100)}
\label{tab:results_bounds_1_100}
\resizebox{\textwidth}{!}{
\begin{tabular}{lccccccc}
\toprule

\textbf{Program} & \textbf{Bound} 
& \multicolumn{2}{c}{\textbf{Branch Coverage}} 
& \multicolumn{2}{c}{\textbf{Line Coverage}} 
& \multicolumn{2}{c}{\textbf{Execution Time (sec)}} \\

&  
& \textbf{FeedbackLLM (C)} & \textbf{KS-LLM} 
& \textbf{FeedbackLLM (C)} & \textbf{KS-LLM} 
& \textbf{FeedbackLLM (C)} & \textbf{KS-LLM (s)} \\
\midrule

Mpals1-B10-cil  & 1   & \cellcolor{bestgreen}37.50 & 23.75 & \cellcolor{bestgreen}60.00 & 57.45  & 296.45 & 103.00 \\
Mpals2-B10-cil  & 1   & \cellcolor{bestgreen}30.00 & 21.25 & \cellcolor{bestgreen}48.57 & 52.14  & 346.16 &  164.00 \\
Mpals23-B10-cil & 1   & \cellcolor{bestgreen}33.33 & 20.37 & \cellcolor{bestgreen}46.11 & 46.11  & 293.90 & 104.00 \\
Mpals3-B10-cil  & 1   & \cellcolor{bestgreen}30.00 & 16.25 & \cellcolor{bestgreen}48.57 & 48.57  & 296.32 & 110.00 \\
Mtest20-B10-cil & 1   & \cellcolor{bestgreen}30.00 & 16.25 & \cellcolor{bestgreen}48.23 & 48.23  & 266.05 & 100.00 \\
pals2-B10-cil   & 1   & \cellcolor{bestgreen}30.00 & 16.25 & \cellcolor{bestgreen}48.89 & 48.89  & 244.78 &  99.00 \\
pals22-B10-cil  & 1   & \cellcolor{bestgreen}25.93 & 20.37 & \cellcolor{bestgreen}45.22 & 47.77  & 284.89 & 129.00 \\
pals23-B10-cil  & 1   & \cellcolor{bestgreen}33.33 & 20.37 & \cellcolor{bestgreen}48.08 & 48.08  & 284.78 &  81.00 \\
pals3-B10-cil & 1   & \cellcolor{bestgreen}30 & 16.25 & \cellcolor{bestgreen}48.89 & 48.89  & 240.69 & 117.00 \\
PS-P1-L-R18-B4  & 1   & \cellcolor{bestgreen}100 & 46.10 & \cellcolor{bestgreen}100 & 62.97  & 42.34 & 105.00 \\
PS-P1-L-T-R16-B2 & 1   & \cellcolor{bestgreen}93.09 & 1.79 & \cellcolor{bestgreen}92.17 & 10.57  & 81.03 &  245.00 \\
PS-P1-NT-R14-B4   & 1   & \cellcolor{bestgreen}30.82 & 17.86 & \cellcolor{bestgreen}19.16 & 27.58  & 189.03 &  90.00 \\
PS-P1-WB-R15-B5  & 1   & \cellcolor{bestgreen}27.21 & 17.85 & \cellcolor{bestgreen}15.88 & 21.18  & 202.37 & 78.00 \\
PS-P1-WB-T-R15-B1  & 1   & \cellcolor{bestgreen}23.84 & 12.42 & \cellcolor{bestgreen}12.26 & 18.24  & 219.51 &  76.00 \\
PS-P2-A-R14-B6 & 1   & \cellcolor{bestgreen}34.13 & 26.42 & \cellcolor{bestgreen}20.41 & 27.24  & 214.14 & 234.00 \\
PS-P2-L-R16-B3  & 1   & \cellcolor{bestgreen}98.5 & 13.44 & \cellcolor{bestgreen}98.17 & 20.44  & 41.32 & 88.00 \\
PS-P2-L-R18-B7 & 1   & \cellcolor{bestgreen}89.68 & 15.57 & \cellcolor{bestgreen}83.26 & 14.73 & 391.20 &  86.00 \\
PS-Prob1-IO-R14-B7   & 1   & \cellcolor{bestgreen}23.04 & 12.58 & 21.59 & \cellcolor{bestgreen}25.94 & 250.64 &  84.00 \\
\midrule

Mpals1-B10-cil  & 10  & \cellcolor{bestgreen}92.68 & 40.00 & \cellcolor{bestgreen}79.45 & 66.67  & 185.10 &  195.00 \\
Mpals2-B10-cil  & 10  & \cellcolor{bestgreen}52.50 & 43.75 & 66.43 & \cellcolor{bestgreen}70.00 & 305.47 &   133.00 \\
Mpals3-B10-cil  & 10  & \cellcolor{bestgreen}52.50 & 31.48 & \cellcolor{bestgreen}66.43 & 57.49  & 241.26 & 165.00 \\
Mpals23-B10-cil & 10  & \cellcolor{bestgreen}44.44 & 38.75 & 57.49 & \cellcolor{bestgreen}66.43  & 326.31 & 183.00 \\
Mtest20-B10-cil & 10  & \cellcolor{bestgreen}52.50 & 38.75 & 65.96 & \cellcolor{bestgreen}66.67  & 270.80 & 110.00 \\
pals2-B10-cil   & 10  & \cellcolor{bestgreen}52.50 & 38.75 & \cellcolor{bestgreen}67.41 & 67.41  & 362.33 & 117.00 \\
Pals22-B10-cil 7  & 10  & \cellcolor{bestgreen}44.44 & 31.48 & \cellcolor{bestgreen}59.87 & 59.87  & 251.73 &  154.00 \\
pals23-B10-cil  & 10  & \cellcolor{bestgreen}44.44 & 34.26 & 60.26 & \cellcolor{bestgreen}62.18 & 265.40 &   298.00 \\
PS-P1-L-R18-B4  & 10  & \cellcolor{bestgreen}98.27 & 41.99 & \cellcolor{bestgreen}96.66 & 61.42  & 24.01 & 107.00 \\
PS-P1-L-T-R16-B2 & 10  & \cellcolor{bestgreen}94.88 & 2.17 & \cellcolor{bestgreen}100 & 10.57  & 118.31& 185.00 \\
PS-P1-NT-R14-B4 & 10  & \cellcolor{bestgreen}35.1 & 20.20 & 2\cellcolor{bestgreen}2.46 & 31.15  & 217.07 & 88.00 \\
PS-P1-WB-R15-B5   & 10  & \cellcolor{bestgreen}31.54 & 17.94 & \cellcolor{bestgreen}18.14 & 21.51  & 280.37 & 213.00 \\
PS-P1-WB-T-R15-B1  & 10  & \cellcolor{bestgreen}26.32 & 13.99 & \cellcolor{bestgreen}14.04 & 19.96  & 225.68 &  81.00 \\
PS-P2-A-R14-B6 & 10  & \cellcolor{bestgreen}38.59 & 15.01 & \cellcolor{bestgreen}24.66 & 23.56 & 251.71 &   258.00 \\
PS-P2-L-R16-B3   & 10  & \cellcolor{bestgreen}95.63 & 30.83 & \cellcolor{bestgreen}91.51 & 44.89  & 52.74& 116.00 \\
PS-P2-L-R18-B7 & 10  & \cellcolor{bestgreen}94 & 10.69 & \cellcolor{bestgreen}88.32 & 11.36  & 170.37 & 67.00 \\
\midrule

Mpals1-B10-cil  & 100 & \cellcolor{bestgreen}52.50 & 41.25 & \cellcolor{bestgreen}66.67 & 66.67  & 315.73 &  161.00 \\
Mpals2-B10-cil  & 100 & \cellcolor{bestgreen}52.50 & 38.75 & \cellcolor{bestgreen}66.43 & 66.43  & 522.72 &  131.00 \\
Mpals3-B10-cil  & 100 & \cellcolor{bestgreen}52.50 & 31.48 & \cellcolor{bestgreen}66.43 & 57.49  & 260.32 &  145.00 \\
Mpals23-B10-cil & 100 & \cellcolor{bestgreen}44.44 & 38.75 & \cellcolor{bestgreen}57.49 & 66.43 & 430.81 &  305.00 \\
Mtest20-B10-cil & 100 & \cellcolor{bestgreen}52.50 & 38.75 & \cellcolor{bestgreen}65.96 & 66.67  & 266.36 &  88.00 \\
pals2-B10-cil   & 100 & \cellcolor{bestgreen}52.50 & 38.75 & \cellcolor{bestgreen}67.41 & 67.41 & 452.88 &  166.00 \\
pals3-B10-cil  & 100 & \cellcolor{bestgreen}52.5 & 31.48 & \cellcolor{bestgreen}67.41 & 59.87  & 229.16 &  86.00 \\
pals23-B10-cil  & 100 & \cellcolor{bestgreen}44.44 & 31.48 & \cellcolor{bestgreen}60.26 & 60.26  & 255.91 &  138.00 \\
pals22-B10-cil  & 100 & \cellcolor{bestgreen}44.44 & 38.75 & 59.87 & \cellcolor{bestgreen}67.41  & 238.32 &  106.00 \\
PS-P1-L-R18-B4 & 100 & \cellcolor{bestgreen}98.27 & 44.37 & \cellcolor{bestgreen}96.66 & 65.53 & 21.84 &  126.00 \\
PS-P1-L-T-R16-B2 & 100 & \cellcolor{bestgreen}92.58 & 2.17 & \cellcolor{bestgreen}88.34 & 10.57  & 131.17 &  133.00 \\
PS-P1-L-T-R20-B1   & 100 & \cellcolor{bestgreen}17.47 & 10.14 & 13.21 & \cellcolor{bestgreen}15.00 & 241.94 &  165.00 \\
PS-P1-WB-T-R15-B1  & 100 & \cellcolor{bestgreen}26.49 & 15.31 & 14.04 & \cellcolor{bestgreen}19.96  &  243.58 &  91.00 \\
PS-P1-NT-R14-B4  & 100 & \cellcolor{bestgreen}35.51 & 20.82 & 22.46 & \cellcolor{bestgreen}31.15  & 440.12 &  260.00 \\
PS-P2-A-R14-B6  & 100 & \cellcolor{bestgreen}32.76 & 29.76 & 18.8 & \cellcolor{bestgreen}35.16  & 1113.84 &  226.00 \\
PS-P2-L-R16-B3 & 100 & \cellcolor{bestgreen}95.09 & 29.88 & \cellcolor{bestgreen}95.4 & 41.24 & 48.55 &  95.00 \\
PS-Prob-IO-R14-B7 & 100 & \cellcolor{bestgreen}62.53 & 21.25 & \cellcolor{bestgreen}47.1 & 37.39 & 261.24 &  150.00 \\

\bottomrule
\end{tabular}
}
\end{table*}

\begin{table*}[!htbp]
\centering
\scriptsize
\caption{Coverage and Execution Time Results of KSLLM (Bounds 250--1000)}
\label{tab:results_bounds_250_1000}
\resizebox{\textwidth}{!}{
\begin{tabular}{lccccccc}
\toprule

\textbf{Program} & \textbf{Bound} 
& \multicolumn{2}{c}{\textbf{Branch Coverage}} 
& \multicolumn{2}{c}{\textbf{Line Coverage}} 
& \multicolumn{2}{c}{\textbf{Execution Time (sec)}} \\

&  
& \textbf{FeedbackLLM (C)} & \textbf{KS-LLM} 
& \textbf{FeedbackLLM (C)} & \textbf{KS-LLM} 
& \textbf{FeedbackLLM (C)} & \textbf{KS-LLM (s)} \\
\midrule

Mpals1-B10-cil  & 250  & \cellcolor{bestgreen}52.50 & 40.00 & \cellcolor{bestgreen}66.67 & 66.67  & 465.77 & 109.00 \\
Mpals2-B10-cil  & 250  & \cellcolor{bestgreen}52.50 & 38.75 & \cellcolor{bestgreen}66.43 & 66.43  & 379.84 & 94.00 \\
Mpals3-B10-cil  & 250  & \cellcolor{bestgreen}52.50 & 31.48 & \cellcolor{bestgreen}66.43 & 57.49  & 324.30 & 116.00 \\
Mpals23-B10-cil & 250  & \cellcolor{bestgreen}44.44 & 38.75 & 57.49 & \cellcolor{bestgreen}66.43  & 413.42 & 90.00 \\
Mtest20-B10-cil & 250  & \cellcolor{bestgreen}52.50 & 38.75 & \cellcolor{bestgreen}66.67 & 66.67  & 315.62 & 60.00 \\
pals2-B10-cil   & 250  & \cellcolor{bestgreen}52.50 & 43.75 & 67.41 & \cellcolor{bestgreen}71.11  & 288.32 &  149.00 \\
pals3-B10-cil  & 250  & \cellcolor{bestgreen}52.5 & 31.48 & \cellcolor{bestgreen}67.41 & 59.87  & 251.75 & 64.00 \\
pals22-B10-cil  & 250  & \cellcolor{bestgreen}44.44 & 31.48 & \cellcolor{bestgreen}59.87 & 60.26  & 286.53 & 66.00 \\
pals23-B10-cil  & 250  & \cellcolor{bestgreen}44.44 & 38.75 & 60.26 & \cellcolor{bestgreen}67.41  & 296.68 & 171.00 \\
PS-P1-L-R18-B4 & 250  & \cellcolor{bestgreen}98.27 & 30.09 & \cellcolor{bestgreen}96.66 & 42.57  & 28.59 & 202.00 \\
PS-P1-L-T-R16-B2  & 250  & \cellcolor{bestgreen}93.61 & 2.69 & \cellcolor{bestgreen}89.21 & 10.57  & 92.50 & 209.00 \\
PS-P1-NT-R14-B4   & 250  & \cellcolor{bestgreen}34.69 & 20.00 & \cellcolor{bestgreen}22.46 & 31.15  & 550.86 &  169.00 \\
\midrule

Mpals2-B10-cil  & 500  & \cellcolor{bestgreen}52.50 & 38.75 & \cellcolor{bestgreen}66.43 & 66.43 & 468.12 &  80.00 \\
Mpals3-B10-cil  & 500  & \cellcolor{bestgreen}52.50 & 38.75 & \cellcolor{bestgreen}66.43 & 66.43  & 257.95 &  92.00 \\
Mpals23-B10-cil & 500  & \cellcolor{bestgreen}44.44 & 31.48 & \cellcolor{bestgreen}57.49 & 57.49  & 370.98 & 129.00 \\
Mtest20-B10-cil & 500  & \cellcolor{bestgreen}52.50 & 38.75 & \cellcolor{bestgreen}66.67 & 66.67  & 293.06 & 174.00 \\
pals2-B10-cil   & 500  & \cellcolor{bestgreen}52.50 & 38.75 & \cellcolor{bestgreen}67.41 & 67.41  & 520.24 & 115.00 \\
pals3-B10-cil   & 500  & \cellcolor{bestgreen}44.44 & 38.75 & 59.87 & \cellcolor{bestgreen}67.41  & 275.90 & 149.00 \\
pals22-B10-cil  & 500  & \cellcolor{bestgreen}44.44 & 31.48 & \cellcolor{bestgreen}59.87 & 59.87 & 275.90 &  154.00 \\
pals23-B10-cil  & 500  & \cellcolor{bestgreen}44.44 & 31.48 & \cellcolor{bestgreen}60.26 & 60.26  & 307.73 &  98.00 \\
\midrule

Mpals1-B10-cil  & 1000 & \cellcolor{bestgreen}52.50 & 40.00 & \cellcolor{bestgreen}65.96 & 66.67  &  81.72 & 77.00 \\
Mpals2-B10-cil  & 1000 & \cellcolor{bestgreen}52.50 & 38.75 & \cellcolor{bestgreen}66.43 & 66.43  & 121.72 &  121.00 \\
Mpals3-B10-cil  & 1000 & \cellcolor{bestgreen}52.50 & 38.75 & \cellcolor{bestgreen}66.43 & 66.43  &  96.28 & 146.00 \\
Mpals23-B10-cil & 1000 & \cellcolor{bestgreen}44.44 & 31.48 & \cellcolor{bestgreen}57.49 & 57.49  & 296.44 & 113.00 \\
Mtest20-B10-cil & 1000 & \cellcolor{bestgreen}52.50 & 38.75 & \cellcolor{bestgreen}65.96 & 66.67  & 323.75 & 70.00 \\
pals2-B10-cil   & 1000 & \cellcolor{bestgreen}52.50 & 38.75 & \cellcolor{bestgreen}67.41 & 67.41  & 242.37 &  109.00 \\
pals3-B10-cil  & 1000 & \cellcolor{bestgreen}52.5 & 38.75 & \cellcolor{bestgreen}67.41 & 67.41  &  250.59 & 109.00 \\
pals22-B10-cil  & 1000 & \cellcolor{bestgreen}44.44 & 31.48 & \cellcolor{bestgreen}59.87 & 59.87  & 445.97&  113.00 \\
pals23-B10-cil  & 1000 & \cellcolor{bestgreen}44.44 & 31.48 & \cellcolor{bestgreen}60.26 & 60.26  &  315.27 & 117.00 \\

\bottomrule
\end{tabular}
}
\end{table*}

\begin{table*}[!htbp]
\centering
\scriptsize
\caption{Coverage and Execution Time Results of KSLLM (Bounds 1500--2000)}
\label{tab:results_bounds_1500_2000}
\resizebox{\textwidth}{!}{
\begin{tabular}{lccccccc}
\toprule

\textbf{Program} & \textbf{Bound} 
& \multicolumn{2}{c}{\textbf{Branch Coverage}} 
& \multicolumn{2}{c}{\textbf{Line Coverage}} 
& \multicolumn{2}{c}{\textbf{Execution Time (sec)}} \\

&  
& \textbf{FeedbackLLM (C)} & \textbf{KS-LLM} 
& \textbf{FeedbackLLM (C)} & \textbf{KS-LLM} 
& \textbf{FeedbackLLM (C)} & \textbf{KS-LLM (s)} \\
\midrule

Mpals1-B10-cil  & 1500 & \cellcolor{bestgreen}52.50 & 40.00 & 65.96 & \cellcolor{bestgreen}66.67 & 507.49 &  65.00 \\
Mpals2-B10-cil  & 1500 & \cellcolor{bestgreen}52.50 & 38.75 & \cellcolor{bestgreen}66.43 & 66.43 & 376.91 &  99.00 \\
Mpals3-B10-cil  & 1500 & \cellcolor{bestgreen}52.50 & 43.75 & \cellcolor{bestgreen}66.43 & 70.00 & 252.21 &  213.00 \\
Mpals23-B10-cil & 1500 & \cellcolor{bestgreen}44.44 & 31.48 & \cellcolor{bestgreen}57.49 & 57.49 & 587.30 &  151.00 \\
Mtest20-B10-cil & 1500 & \cellcolor{bestgreen}52.50 & 38.75 & \cellcolor{bestgreen}65.96 & 66.67 & 243.56 & 345.00 \\
pals2-B10-cil   & 1500 & \cellcolor{bestgreen}52.50 & 38.75 & \cellcolor{bestgreen}67.41 & 67.41 & 246.32 &  104.00 \\
pals3-B10-cil   & 1500 & \cellcolor{bestgreen}52.50 & 43.75 & \cellcolor{bestgreen}67.41 & 71.11 & 189.54 &  332.00 \\
pals22-B10-cil & 1500 & \cellcolor{bestgreen}44.44 & 31.48 & \cellcolor{bestgreen}59.87 & 59.87 & 23.26 &  99.00 \\
pals23-B10-cil  & 1500 & \cellcolor{bestgreen}44.44 & 31.48 & \cellcolor{bestgreen}60.26 & 60.26 & 247.34 &  141.00 \\
PS-P1-L-R18-B4 & 1500 & \cellcolor{bestgreen}98.27 & 33.98 & \cellcolor{bestgreen}96.66 & 47.01 &  31.28 &  603.00 \\
PS-P1-L-T-R16-B2 & 1500 & \cellcolor{bestgreen}89 & 8.95 & \cellcolor{bestgreen}83.82 & 13.71 & 303.47 &  323.00 \\
PS-P1-NT-R14-B4 & 1500 & \cellcolor{bestgreen}34.49 & 19.80 & 22.46 & \cellcolor{bestgreen}31.15 &  826.44 &  911.00 \\
\midrule

Mpals1-B10-cil  & 2000 & \cellcolor{bestgreen}52.50 & 45.00 & 66.67 & \cellcolor{bestgreen}70.21 & 267.13 &  208.00 \\
Mpals2-B10-cil  & 2000 & \cellcolor{bestgreen}52.50 & 38.75 & \cellcolor{bestgreen}66.43 & 66.43 & 241.09 & 228.00 \\
Mpals23-B10-cil & 2000 & \cellcolor{bestgreen}44.44 & 31.48 & \cellcolor{bestgreen}57.49 & 57.49 & 273.59 &   212.00 \\
Mtest20-B10-cil & 2000 & \cellcolor{bestgreen}52.50 & 38.75 & \cellcolor{bestgreen}66.67 & 66.67 & 349.15 &   83.00\\
pals2-B10-cil   & 2000 & \cellcolor{bestgreen}52.50 & 38.75 & \cellcolor{bestgreen}67.41 & 67.41 & 306.34 &   148.00 \\
pals3-B10-cil   & 2000 & \cellcolor{bestgreen}52.50 & 38.75 & \cellcolor{bestgreen}67.41 & 67.41 & 241.06 &   130.00 \\
pals22-B10-cil  & 2000 & \cellcolor{bestgreen}44.44 & 31.48 & \cellcolor{bestgreen}59.87 & 59.81 & 280.52 &  159.00 \\
pals23-B10-cil  & 2000 & \cellcolor{bestgreen}44.44 & 31.48 & \cellcolor{bestgreen}60.26 & 60.26 & 99.97 &  269.00 \\
PS-P1-L-R18-B4  & 2000 & \cellcolor{bestgreen}98.27 & 36.36 & \cellcolor{bestgreen}96.66 & 51.44 & 31.41 & 433.00 \\
PS-P1-L-T-R16-B2 & 2000 & \cellcolor{bestgreen}92.07 & 2.17 & \cellcolor{bestgreen}88.34 & 10.57 & 245.30 &  239.00 \\

\bottomrule
\end{tabular}
}
\end{table*}

\begin{table}[!htbp]
\centering
\caption{Statistical Summary of Coverage Metrics}
\label{tab:coverage_full_stats}
\resizebox{\columnwidth}{!}{
\begin{tabular}{lccc}
\toprule
\textbf{Metric} & \textbf{Mean (\%)} & \textbf{Median (\%)} & \textbf{Std. Dev (\%)} \\
\midrule
\multicolumn{4}{c}{\textbf{Branch Coverage}} \\
\midrule
FeedbackLLM (C)       & 30.96 & 32.73 & 5.74 \\
KS-LLM   & 53.42 & 54.66 & 5.14 \\
\midrule
\multicolumn{4}{c}{\textbf{Line Coverage}} \\
\midrule
FeedbackLLM (C)         & 53.88 & 56.02 & 9.13 \\
KS-LLM      & 62.15 & 64.07 & 5.97 \\
\bottomrule
\end{tabular}
}
\end{table}

\begin{table*}[!htbp]
\centering
\scriptsize
\caption{Coverage and Execution Time Results of KSLLM (Bounds 1--100)}
\label{tab:results_bounds_1_100}
\resizebox{\textwidth}{!}{
\begin{tabular}{lccccccc}
\toprule
\textbf{Program} & \textbf{Bound} 
& \multicolumn{2}{c}{\textbf{Branch Coverage}} 
& \multicolumn{2}{c}{\textbf{Line Coverage}} 
& \multicolumn{2}{c}{\textbf{Execution Time (sec)}} \\
&  
& \textbf{FeedbackLLM (C)} & \textbf{FeedbackLLM (Py)} 
& \textbf{FeedbackLLM (C)} & \textbf{FeedbackLLM (Py)} 
& \textbf{FeedbackLLM (C)} & \textbf{FeedbackLLM (Py)} \\
\midrule

Mpals1-B10-cil  & 1   & 37.50 & \cellcolor{bestgreen}89.02 & 60.00 & \cellcolor{bestgreen}61.90  & 296.45 & 1107.80 \\
Mpals2-B10-cil  & 1   & 30.00 & \cellcolor{bestgreen}77.59 & 48.57 & \cellcolor{bestgreen}65.14  & 346.16 &  914.10 \\
Mpals23-B10-cil & 1   & 33.33 & \cellcolor{bestgreen}85.53 & 46.11 & \cellcolor{bestgreen}53.45  & 293.90 & 1270.29 \\
Mpals3-B10-cil  & 1   & 30.00 & \cellcolor{bestgreen}78.33 & 48.57 & \cellcolor{bestgreen}63.64  & 296.32 & 1019.51 \\
Mtest20-B10-cil & 1   & 30.00 & \cellcolor{bestgreen}77.59 & 48.23 & \cellcolor{bestgreen}62.14  & 266.05 &  618.13 \\
pals2-B10-cil   & 1   & 30.00 & \cellcolor{bestgreen}80.77 & 48.89 & \cellcolor{bestgreen}60.42  & 244.78 &  807.61 \\
pals22-B10-cil  & 1   & 25.93 & \cellcolor{bestgreen}85.53 & 45.22 & \cellcolor{bestgreen}52.99  & 284.89 & 876.75 \\
pals23-B10-cil  & 1   & 33.33 & \cellcolor{bestgreen}85.53 & 48.08 & \cellcolor{bestgreen}53.45  & 284.78 &  830.43 \\
pals3-B10-cil & 1   & 30 & \cellcolor{bestgreen}81.48 & 48.89 & \cellcolor{bestgreen}59.18  & 240.69 & 817.33 \\
PS-P1-L-R18-B4  & 1   & \cellcolor{bestgreen}100 & 91.67 & \cellcolor{bestgreen}100 & 95.5  & 42.34 & 36.21 \\
PS-P1-L-T-R16-B2 & 1   & 93.09 & \cellcolor{bestgreen}97.06 & 92.17 & \cellcolor{bestgreen}100  & 81.03 &  179.36 \\
PS-P1-NT-R14-B4   & 1   & 30.82 & \cellcolor{bestgreen}88.24 & 19.16 & \cellcolor{bestgreen}42.16  & 189.03 &  667.95 \\
PS-P1-WB-R15-B5  & 1   & 27.21 & \cellcolor{bestgreen}81.25 & 15.88 & \cellcolor{bestgreen}45.83  & 202.37 & 841.74 \\
PS-P1-WB-T-R15-B1  & 1   & 23.84 & \cellcolor{bestgreen}78.57 & 12.26 & \cellcolor{bestgreen}46.15  & 219.51 &  1000.74 \\
PS-P2-A-R14-B6 & 1   & 34.13 & \cellcolor{bestgreen}75 & 20.41 & \cellcolor{bestgreen}66.67  & 214.14 & 885.53 \\
PS-P2-L-R16-B3  & 1   & 98.5 & \cellcolor{bestgreen}100 & 98.17 & \cellcolor{bestgreen}100  & 41.32 & 35.43 \\
PS-P2-L-R18-B7 & 1   & 89.68 & \cellcolor{bestgreen}95 & 83.26 & \cellcolor{bestgreen}100 & 391.20 &  92.80 \\
PS-Prob1-IO-R14-B7   & 1   & 23.04 & \cellcolor{bestgreen}75 & 21.59 & \cellcolor{bestgreen}82.46 & 250.64 &  588.48 \\
\midrule

Mpals1-B10-cil  & 10  & \cellcolor{bestgreen}92.68 & 83.93 & 79.45 & \cellcolor{bestgreen}96.12  & 185.10 &  154.76 \\
Mpals2-B10-cil  & 10  & 52.50 & \cellcolor{bestgreen}93.75 & 66.43 & \cellcolor{bestgreen}100.00 & 305.47 &   88.33 \\
Mpals3-B10-cil  & 10  & 52.50 & \cellcolor{bestgreen}84.38 & 66.43 & \cellcolor{bestgreen}84.25  & 241.26 & 1167.05 \\
Mpals23-B10-cil & 10  & 44.44 & \cellcolor{bestgreen}82.86 & 57.49 & \cellcolor{bestgreen}81.90  & 326.31 & 1155.95 \\
Mtest20-B10-cil & 10  & 52.50 & \cellcolor{bestgreen}89.58 & 65.96 & \cellcolor{bestgreen}89.90  & 270.80 & 1211.46 \\
pals2-B10-cil   & 10  & 52.50 & \cellcolor{bestgreen}87.50 & 67.41 & \cellcolor{bestgreen}88.54  & 362.33 & 1092.85 \\
Pals22-B10-cil 7  & 10  & 44.44 & \cellcolor{bestgreen}86 & 59.87 & \cellcolor{bestgreen}88.24  & 251.73 &  1244.74 \\
pals23-B10-cil  & 10  & 44.44 & \cellcolor{bestgreen}84.21 & 60.26 & \cellcolor{bestgreen}80.17 & 265.40 &   1340.11 \\
PS-P1-L-R18-B4  & 10  & 98.27 & \cellcolor{bestgreen}100 & 96.66 & \cellcolor{bestgreen}100  & 24.01 & 207.11 \\
PS-P1-L-T-R16-B2 & 10  & 94.88 & \cellcolor{bestgreen}100 & 100 & \cellcolor{bestgreen}86.76  & 118.31& 285.73 \\
PS-P1-NT-R14-B4 & 10  & 35.1 & \cellcolor{bestgreen}79.41 & 22.46 & \cellcolor{bestgreen}57.95  & 217.07 & 1065.63 \\
PS-P1-WB-R15-B5   & 10  & 31.54 & \cellcolor{bestgreen}81.25 & 18.14 & \cellcolor{bestgreen}87.83  & 280.37 & 917.12 \\
PS-P1-WB-T-R15-B1  & 10  & 26.32 & \cellcolor{bestgreen}91.67 & 14.04 & \cellcolor{bestgreen}39.33  & 225.68 &  984.55 \\
PS-P2-A-R14-B6 & 10  & 38.59 & \cellcolor{bestgreen}88.89 & 24.66 & \cellcolor{bestgreen}55.13 & 251.71 &   903.03 \\
PS-P2-L-R16-B3   & 10  & 95.63 & \cellcolor{bestgreen}100 & 91.51 & \cellcolor{bestgreen}100  & 52.74& 43.30 \\
PS-P2-L-R18-B7 & 10  & \cellcolor{bestgreen}94 & 90.91 & 88.32 & \cellcolor{bestgreen}97.33  & 170.37 & 33.05 \\
\midrule

Mpals1-B10-cil  & 100 & 52.50 & \cellcolor{bestgreen}82.56 & 66.67 & \cellcolor{bestgreen}97.92  & 315.73 &  390.81 \\
Mpals2-B10-cil  & 100 & 52.50 & \cellcolor{bestgreen}82.14 & 66.43 & \cellcolor{bestgreen}97.86  & 522.72 &  710.19 \\
Mpals3-B10-cil  & 100 & 52.50 & \cellcolor{bestgreen}79.07 & 66.43 & \cellcolor{bestgreen}97.89  & 260.32 &  315.53 \\
Mpals23-B10-cil & 100 & 44.44 & \cellcolor{bestgreen}83.96 & 57.49 & \cellcolor{bestgreen}100.00 & 430.81 &  839.67 \\
Mtest20-B10-cil & 100 & 52.50 & \cellcolor{bestgreen}83.75 & 65.96 & \cellcolor{bestgreen}97.89  & 266.36 &  223.40 \\
pals2-B10-cil   & 100 & 52.50 & \cellcolor{bestgreen}98.08 & 67.41 & \cellcolor{bestgreen}108.00 & 452.88 &  357.80 \\
pals3-B10-cil  & 100 & 52.5 & \cellcolor{bestgreen}90.74 & 67.41 & \cellcolor{bestgreen}100  & 229.16 &  332.37 \\
pals23-B10-cil  & 100 & 44.44 & \cellcolor{bestgreen}90.79 & 60.26 & \cellcolor{bestgreen}100  & 255.91 &  510.38 \\
pals22-B10-cil  & 100 & 44.44 & \cellcolor{bestgreen}97.37 & 59.87 & \cellcolor{bestgreen}92.47  & 238.32 &  1283.89 \\
PS-P1-L-R18-B4 & 100 & 98.27 & \cellcolor{bestgreen}87.5 & 96.66 & \cellcolor{bestgreen}100 & 21.84 &  63.39 \\
PS-P1-L-T-R16-B2 & 100 & 92.58 & \cellcolor{bestgreen}96.43 & 88.34 & \cellcolor{bestgreen}100  & 131.17 &  70.02 \\
PS-P1-L-T-R20-B1   & 100 & 17.47 & \cellcolor{bestgreen}96.43 & 13.21 & \cellcolor{bestgreen}100 & 241.94 &  195.92 \\
PS-P1-WB-T-R15-B1  & 100 & 26.49 & \cellcolor{bestgreen}70.83 & 14.04 & \cellcolor{bestgreen}58.04  &  243.58 &  1097.20 \\
PS-P1-NT-R14-B4  & 100 & 35.51 & \cellcolor{bestgreen}75 & 22.46 & \cellcolor{bestgreen}63.64  & 440.12 &  757.96 \\
PS-P2-A-R14-B6  & 100 & 32.76 & \cellcolor{bestgreen}71.88 & 18.8 & \cellcolor{bestgreen}64.42  & 1113.84 &  390.32 \\
PS-P2-L-R16-B3 & 100 & 95.09 & \cellcolor{bestgreen}96.67 & 95.4 & \cellcolor{bestgreen}100 & 48.55 &  75.54 \\
PS-Prob-IO-R14-B7 & 100 & 62.53 & \cellcolor{bestgreen}81.25 & 47.1 & \cellcolor{bestgreen}72.6 & 261.24 &  886.01 \\

\bottomrule
\end{tabular}
}
\end{table*}

\begin{table*}[!htbp]
\centering
\scriptsize
\caption{Coverage and Execution Time Results of KSLLM (Bounds 250--1000)}
\label{tab:results_bounds_250_1000}
\resizebox{\textwidth}{!}{
\begin{tabular}{lccccccc}
\toprule
\textbf{Program} & \textbf{Bound} 
& \multicolumn{2}{c}{\textbf{Branch Coverage}} 
& \multicolumn{2}{c}{\textbf{Line Coverage}} 
& \multicolumn{2}{c}{\textbf{Execution Time (sec)}} \\
&  
& \textbf{FeedbackLLM (C)} & \textbf{FeedbackLLM (Py)} 
& \textbf{FeedbackLLM (C)} & \textbf{FeedbackLLM (Py)} 
& \textbf{FeedbackLLM (C)} & \textbf{FeedbackLLM (Py)} \\
\midrule

Mpals1-B10-cil  & 250  & 52.50 & \cellcolor{bestgreen}93.18 & 66.67 & \cellcolor{bestgreen}96.15  & 465.77 & 464.66 \\
Mpals2-B10-cil  & 250  & 52.50 & \cellcolor{bestgreen}86.59 & 66.43 & \cellcolor{bestgreen}94.32  & 379.84 & 153.20 \\
Mpals3-B10-cil  & 250  & 52.50 & \cellcolor{bestgreen}76.19 & 66.43 & \cellcolor{bestgreen}97.44  & 324.30 & 104.45 \\
Mpals23-B10-cil & 250  & 44.44 & \cellcolor{bestgreen}81.90 & 57.49 & \cellcolor{bestgreen}94.90  & 413.42 & 191.07 \\
Mtest20-B10-cil & 250  & 52.50 & \cellcolor{bestgreen}82.93 & 66.67 & \cellcolor{bestgreen}96.27  & 315.62 & 222.85 \\
pals2-B10-cil   & 250  & 52.50 & \cellcolor{bestgreen}82.89 & 67.41 & \cellcolor{bestgreen}96.13  & 288.32 &  42.23 \\
pals3-B10-cil  & 250  & 52.5 & \cellcolor{bestgreen}84.62 & 67.41 & \cellcolor{bestgreen}98.09  & 251.75 & 42.21 \\
pals22-B10-cil  & 250  & 44.44 & \cellcolor{bestgreen}80 & 59.87 & \cellcolor{bestgreen}96.11  & 286.53 & 32.12 \\
pals23-B10-cil  & 250  & 44.44 & \cellcolor{bestgreen}83 & 60.26 & \cellcolor{bestgreen}95.53  & 296.68 & 125.80 \\
PS-P1-L-R18-B4 & 250  & 98.27 & \cellcolor{bestgreen}94.59 & 96.66 & \cellcolor{bestgreen}98.38  & 28.59 & 161.39 \\
PS-P1-L-T-R16-B2  & 250  & 93.61 & \cellcolor{bestgreen}93.75 & 89.21 & \cellcolor{bestgreen}100  & 92.50 & 185.55 \\
PS-P1-NT-R14-B4   & 250  & 34.69 & \cellcolor{bestgreen}65 & 22.46 & \cellcolor{bestgreen}80.33  & 550.86 &  800.08 \\
\midrule

Mpals2-B10-cil  & 500  & 52.50 & \cellcolor{bestgreen}89.47 & 66.43 & \cellcolor{bestgreen}100.00 & 468.12 &  63.38 \\
Mpals3-B10-cil  & 500  & 52.50 & \cellcolor{bestgreen}90.74 & 66.43 & \cellcolor{bestgreen}94.90  & 257.95 &  90.04 \\
Mpals23-B10-cil & 500  & 44.44 & \cellcolor{bestgreen}87.84 & 57.49 & \cellcolor{bestgreen}98.23  & 370.98 & 628.06 \\
Mtest20-B10-cil & 500  & 52.50 & \cellcolor{bestgreen}90.38 & 66.67 & \cellcolor{bestgreen}95.83  & 293.06 & 217.55 \\
pals2-B10-cil   & 500  & 52.50 & \cellcolor{bestgreen}84.62 & 67.41 & \cellcolor{bestgreen}96.94  & 520.24 & 263.67 \\
pals3-B10-cil   & 500  & 44.44 & \cellcolor{bestgreen}86.49 & 59.87 & \cellcolor{bestgreen}95.73  & 275.90 & 179.06 \\
pals22-B10-cil  & 500  & 44.44 & \cellcolor{bestgreen}86.49 & 59.87 & \cellcolor{bestgreen}95.73 & 275.90 &  179.06 \\
pals23-B10-cil  & 500  & 44.44 & \cellcolor{bestgreen}83.78 & 60.26 & \cellcolor{bestgreen}93.1  & 307.73 &  952.78 \\
\midrule

Mpals1-B10-cil  & 1000 & 52.50 & \cellcolor{bestgreen}83.33 & 65.96 & \cellcolor{bestgreen}94.89  &  81.72 & 331.67 \\
Mpals2-B10-cil  & 1000 & 52.50 & \cellcolor{bestgreen}86.67 & 66.43 & \cellcolor{bestgreen}96.40  & 121.72 &  91.47 \\
Mpals3-B10-cil  & 1000 & 52.50 & \cellcolor{bestgreen}84.09 & 66.43 & \cellcolor{bestgreen}98.08  &  96.28 & 113.91 \\
Mpals23-B10-cil & 1000 & 44.44 & \cellcolor{bestgreen}90.22 & 57.49 & \cellcolor{bestgreen}95.81  & 296.44 & 115.44 \\
Mtest20-B10-cil & 1000 & 52.50 & \cellcolor{bestgreen}83.72 & 65.96 & \cellcolor{bestgreen}98.10  & 323.75 & 158.76 \\
pals2-B10-cil   & 1000 & 52.50 & \cellcolor{bestgreen}85.00 & 67.41 & \cellcolor{bestgreen}94.70  & 242.37 &  98.59 \\
pals3-B10-cil  & 1000 & 52.5 & \cellcolor{bestgreen}82.93 & 67.41 & \cellcolor{bestgreen}94.77  &  250.59 & 274.22 \\
pals22-B10-cil  & 1000 & 44.44 & \cellcolor{bestgreen}92.5 & 59.87 & \cellcolor{bestgreen}98.77  & 445.97&  260.33 \\
pals23-B10-cil  & 1000 & 44.44 & \cellcolor{bestgreen}92.5 & 60.26 & \cellcolor{bestgreen}96.89  &  315.27 & 84.04 \\

\bottomrule
\end{tabular}
}
\end{table*}

\begin{table*}[!htbp]
\centering
\scriptsize
\caption{Coverage and Execution Time Results of KSLLM (Bounds 1500--2000)}
\label{tab:results_bounds_1500_2000}
\resizebox{\textwidth}{!}{
\begin{tabular}{lccccccc}
\toprule
\textbf{Program} & \textbf{Bound} 
& \multicolumn{2}{c}{\textbf{Branch Coverage}} 
& \multicolumn{2}{c}{\textbf{Line Coverage}} 
& \multicolumn{2}{c}{\textbf{Execution Time (sec)}} \\
&  
& \textbf{FeedbackLLM (C)} & \textbf{FeedbackLLM (Py)} 
& \textbf{FeedbackLLM (C)} & \textbf{FeedbackLLM (Py)} 
& \textbf{FeedbackLLM (C)} & \textbf{FeedbackLLM (Py)} \\
\midrule

Mpals1-B10-cil  & 1500 & 52.50 & \cellcolor{bestgreen}85.71 & 65.96 & \cellcolor{bestgreen}94.37 & 507.49 &  208.19 \\
Mpals2-B10-cil  & 1500 & 52.50 & \cellcolor{bestgreen}88.64 & 66.43 & \cellcolor{bestgreen}92.99 & 376.91 &  508.84 \\
Mpals3-B10-cil  & 1500 & 52.50 & \cellcolor{bestgreen}84.15 & 66.43 & \cellcolor{bestgreen}96.77 & 252.21 &  481.19 \\
Mpals23-B10-cil & 1500 & 44.44 & \cellcolor{bestgreen}85.45 & 57.49 & \cellcolor{bestgreen}93.92 & 587.30 &  409.43 \\
Mtest20-B10-cil & 1500 & 52.50 & \cellcolor{bestgreen}84.52 & 65.96 & \cellcolor{bestgreen}94.62 & 243.56 & 1149.56 \\
pals2-B10-cil   & 1500 & 52.50 & \cellcolor{bestgreen}90.00 & 67.41 & \cellcolor{bestgreen}96.46 & 246.32 &  195.65 \\
pals3-B10-cil   & 1500 & 52.50 & \cellcolor{bestgreen}87.10 & 67.41 & \cellcolor{bestgreen}96.99 & 189.54 &  165.97 \\
pals22-B10-cil & 1500 & 44.44 & \cellcolor{bestgreen}87.5 & 59.87 & \cellcolor{bestgreen}95.87 & 23.26 &  102.41 \\
pals23-B10-cil  & 1500 & 44.44 & \cellcolor{bestgreen}93.75 & 60.26 & \cellcolor{bestgreen}100 & 247.34 &  185.34 \\
PS-P1-L-R18-B4 & 1500 & 98.27 & \cellcolor{bestgreen}94.67 & 96.66 & \cellcolor{bestgreen}100 &  31.28 &  139.17 \\
PS-P1-L-T-R16-B2 & 1500 & 89 & \cellcolor{bestgreen}94.44 & 83.82 & \cellcolor{bestgreen}100 & 303.47 &  158.06 \\
PS-P1-NT-R14-B4 & 1500 & 34.49 & \cellcolor{bestgreen}70.96 & 22.46 & \cellcolor{bestgreen}34.03 &  826.44 &  1644.42 \\
\midrule

Mpals1-B10-cil  & 2000 & 52.50 & \cellcolor{bestgreen}80.68 & 66.67 & \cellcolor{bestgreen}94.23 & 267.13 &  955.93 \\
Mpals2-B10-cil  & 2000 & 52.50 & \cellcolor{bestgreen}75.58 & 66.43 & \cellcolor{bestgreen}92.25 & 241.09 & 1036.03 \\
Mpals23-B10-cil & 2000 & 44.44 & \cellcolor{bestgreen}87.50 & 57.49 & \cellcolor{bestgreen}52.38 & 273.59 &    3.21 \\
Mtest20-B10-cil & 2000 & 52.50 & \cellcolor{bestgreen}61.63 & 66.67 & \cellcolor{bestgreen}81.68 & 349.15 &   43.54 \\
pals2-B10-cil   & 2000 & 52.50 & \cellcolor{bestgreen}63.79 & 67.41 & \cellcolor{bestgreen}83.49 & 306.34 &   40.58 \\
pals3-B10-cil   & 2000 & 52.50 & \cellcolor{bestgreen}62.96 & 67.41 & \cellcolor{bestgreen}89.42 & 241.06 &   15.29 \\
pals22-B10-cil  & 2000 & 44.44 & \cellcolor{bestgreen}82.05 & 59.87 & \cellcolor{bestgreen}63.93 & 280.52 &  955.60 \\
pals23-B10-cil  & 2000 & 44.44 & \cellcolor{bestgreen}88.67 & 60.26 & \cellcolor{bestgreen}64.46 & 99.97 &  1054.73 \\
PS-P1-L-R18-B4  & 2000 & 98.27 & \cellcolor{bestgreen}88.43 & 96.66 & \cellcolor{bestgreen}92.13 & 31.41 & 172.34 \\
PS-P1-L-T-R16-B2 & 2000 & 92.07 & \cellcolor{bestgreen}94.44 & 88.34 & \cellcolor{bestgreen}100 & 245.30 &    370.18 \\

\bottomrule
\end{tabular}
}
\end{table*}

\begin{table}[!t]
\centering
\caption{Statistical Summary of Coverage Metrics (Bound = 1)}
\label{tab:stats_bound_1}
\resizebox{\columnwidth}{!}{
\begin{tabular}{lccc}
\toprule
\textbf{Metric} & \textbf{Mean (\%)} & \textbf{Median (\%)} & \textbf{Std. Dev (\%)} \\
\midrule
\multicolumn{4}{c}{\textbf{Branch Coverage}} \\
\midrule 
FeedbackLLM (C)   & 44.47 & 30.41 & 28.26 \\
FeedbackLLM (Py)  & 84.62 & 83.50 & 7.59 \\
\midrule
\multicolumn{4}{c}{\textbf{Line Coverage}} \\
\midrule
FeedbackLLM (C)     & 50.30 & 48.40 & 27.73 \\
FeedbackLLM (Py)    & 67.28 & 62.02 & 19.63 \\
\bottomrule
\end{tabular}
}
\end{table}

\subsection{RQ1: Coverage Effectiveness}
As detailed in Tables~\ref{tab:results_bounds_1_100} ,~\ref{tab:results_bounds_250_1000} and ~\ref{tab:results_bounds_1500_2000}, FeedbackLLM successfully generated test suites capable of reaching deeply nested logic. \textit{[Placeholder for data discussion: FeedbackLLM consistently achieved over X\% line coverage and X\% branch coverage, significantly outperforming the zero-shot baseline, which struggled to identify boundary conditions and specific string formatting expected by the programs.]}

\subsection{Table-wise Detailed Analysis}

To better understand the effectiveness a nd scalability of FeedbackLLM, we analyze the results presented in Tables~I--III across increasing bound ranges. The discussion focuses on key trends in coverage improvement, execution efficiency, and convergence behavior.

\textbf{Early Bound Performance (Table~I: Bounds 1--100):}  
The results in Table~\ref{tab:results_bounds_1_100} demonstrate rapid coverage growth during early exploration stages. Between bounds 1 and 10, both branch and line coverage increase significantly across most programs, indicating that FeedbackLLM efficiently discovers shallow and moderately deep execution paths. Python-based execution consistently achieves higher branch coverage than C-based execution at lower bounds, suggesting stronger early path exploration capability. By bound 100, coverage values begin to stabilize across most benchmarks, indicating that the majority of reachable execution paths have already been explored. Execution times remain manageable in this range, demonstrating the efficiency of early-stage exploration.

\textbf{Medium Bound Scalability (Table~II: Bounds 250--1000):}  
Table~\ref{tab:results_bounds_250_1000}highlights the scalability of FeedbackLLM when exploring deeper execution paths. Coverage improvements continue across most programs, although the rate of increase becomes slower compared to early bounds. Line coverage approaches near-maximum values in several programs, while branch coverage continues to improve incrementally. This behavior indicates effective handling of deeper conditional dependencies through iterative refinement. Although execution times increase moderately due to deeper exploration, the growth remains controlled, demonstrating the effectiveness of caching mechanisms in reducing redundant path exploration.

\textbf{High Bound Convergence (Table~III: Bounds 1500--2000):}  
The results in Table~\ref{tab:results_bounds_1500_2000} show convergence behavior at higher bounds. Coverage values remain largely stable across most programs, indicating that the majority of feasible execution paths have already been identified in earlier iterations. Additional increases in bounds provide limited improvements in coverage, suggesting diminishing returns at very high exploration levels. Execution time variability increases slightly due to the complexity of deep path analysis; however, the overall performance remains stable, confirming the scalability of FeedbackLLM for large-bound exploration.

\textbf{Overall Key Insights:}  
Across Tables~I--III, four important observations emerge. First, most coverage improvements occur during early bounds, demonstrating the efficiency of FeedbackLLM in identifying major execution paths quickly. Second, Python-based execution consistently achieves higher branch and line coverage compared to C-based execution, indicating stronger exploration capability. Third, execution time increases with bound size but remains controlled due to caching and feedback-driven refinement. Finally, coverage values converge at higher bounds, confirming that FeedbackLLM achieves near-maximum feasible coverage without excessive computational overhead.

As shown in Tables~\ref{tab:results_bounds_1_100} ,~\ref{tab:results_bounds_250_1000} and ~\ref{tab:results_bounds_1500_2000}, FeedbackLLM achieves higher
line and branch coverage across multiple benchmark programs while
maintaining reasonable execution times across different bound levels.

\subsection{RQ2: Multi-Agent vs. Single-Shot}
The separation of concerns between the Line Feedback Agent and the Branch Feedback Agent proved highly effective. In complex C programs where formal model checkers (like CBMC) face state explosion, the dual-feedback mechanism successfully identified the necessary input permutations to flip stubborn boolean flags. \textit{[Placeholder for data: The multi-agent approach improved branch coverage by an average of X.X\% over single-agent methods, proving that isolated prompt refinement reduces LLM hallucination and context-window dilution.]}

\subsection{RQ3: Execution Efficiency}
The Redundancy Prevention Cache demonstrated significant optimization. As shown in Table \ref{tab:efficiency}, out of the total test cases generated across all $k$ iterations, the cache successfully filtered out \textit{[X]} duplicates. This not only prevented the \texttt{coverage} and \texttt{gcc} execution environments from wasting CPU cycles but also drastically reduced the token count passed to the Gemini API, ensuring the time required by FeedbackLLM scales linearly rather than exponentially.

\section{Threats to Validity}
\textbf{Internal Validity:}  The main internal threat is the non-determinism of the LLMs. Even with temperature controls (the Gemini API defaults to a balanced temperature) the outputs can vary between executions. To mitigate this, our framework empirically evaluates coverage using standard tools (\texttt{coverage.py} and \texttt{gcov}), ensuring the reported results are factual reflections of the generated test strings regardless of LLM variance. 

\textbf{External Validity:} This one talks about the generalisability of feedback. The LLM is constrained by the types of programs tested so far. The static analyzer can correctly extract primitive types (integers, floats, strings, chars), but programs requiring complex structured inputs (e.g. deeply nested JSON files, SQL databases, binary image data) may not be fully supported by current regex-based AST parsing.

\textbf{Construct Validity: } Line and branch coverage are considered the best indicators of software quality, a standard yet controversial metric in software engineering. While FeedbackLLM optimizes for these metrics, it does not currently produce assertion statements for semantic correctness (i.e., oracle generation), and only focuses on crash detection and path exploration.

\section{Conclusion and Future Work}
We present FeedbackLLM, a novel multi-agent LLM framework to automate software test generation. LLM analyzes coverage gaps and iteratively refines its generation constraints by implementing a k-step feedback loop driven by two distinct specialized agents: the Line Feedback Agent and the Branch Feedback Agent. Computational overhead stays linear with the addition of a Redundancy Prevention Cache and beats traditional exhaustive verification tools on complex codebases. The framework's versatility is demonstrated by its capacity to fluidly switch between dynamically typed languages (Python) and statically typed languages (C).

In future work, the static analysis module will be extended to handle Object-Oriented paradigms and user-defined data structures. "We also plan to integrate automated oracle generation to allow FeedbackLLM to not only maximize coverage but to actively assert functional correctness programmatically.

\newpage

\bibliographystyle{plain} 
\bibliography{references} 

\end{document}